\documentclass[a4paper,10pt]{article}
\usepackage[utf8]{inputenc}

\title{ Discreteness of time in the evolution of the universe }
\author{Mir Faizal$^1$, Ahmed Farag Ali$^{2}$, Saurya Das$^3$\\ \\
 $^1$Department of Physics and Astronomy, \\ University of Waterloo, \\ Waterloo,
Ontario, N2L 3G1, Canada\\ 
$^2$Department of Physics, Faculty of Sciences,\\  Benha University,  Benha, 13518, Egypt
\\ $^3$Theoretical Physics Group, \\ Department of Physics and Astronomy,\\
University of Lethbridge, 4401 University Drive,\\
Lethbridge, Alberta, Canada T1K 3M4 }
\date{}
\begin{document}

\maketitle

\begin{abstract}
In this paper, we  will first derive the Wheeler-DeWitt equation for the generalized geometry which occurs in M-theory.
Then we will observe that M2-branes act as probes for this generalized geometry, and as M2-branes have an extended structure,
their extended structure will  limits the resolution to which this generalized
geometry can be defined. We will demonstrate that this  will deform  the Wheeler-DeWitt
equation for the generalized geometry. We analyse  such a deformed Wheeler-DeWitt equation in the minisuperspace approximation,
and observe that this deformation can be used as
a solution to the problem of time.  This is because  this deformation    gives rise to time crystals
in our universe due to the spontaneous breaking of time  re-parametrization invariance.
\end{abstract}

\section{Introduction}
It is known that the T–duality is a hidden
symmetry from the spacetime point of view \cite{0j}. However, it is possible to make the
T-duality manifest by doubling the dimensions of space
\cite{1j, 2j, 4j, 5j}.  It has been observed that the
  T–duality is  linearly realized on this geometry, which  turns out to be the generalized geometry
\cite{8j, 9j}. The symmetries in M-theory have also been analysed using the generalized geometry, and in this context
the duality group is related to the U-duality  \cite{  q2, q22, q4}.
The  construction of geometric structures like connection and curvature from a generalized metric has also been analysed
\cite{p2,p4}. The Hamiltonian constraint for M-theory  has also been analysed using  the generalized geometry \cite{geha}.
Furthermore, the duality for M2-branes was also analysed using this construction. However, it is known that M2-branes act as
probes for the generalized geometry in M-theory. Now as M2-branes have an extended structure, the resolution of the generalized
geometry will get limited to the scale fixed by the extended structure of M2-branes. It may be noted that the extended structure
of strings has been used to fix a minimum measurable length scale in perturbative string theory  \cite{z2,zasaqsw,csdcas,cscds,2z}.
This is because
strings are the smallest probes that can be used for analyzing a region of spacetime.  Now just as string act as probes in string theory,
M2-branes act as probes in M-theory. Thus, we do expect such a minimum scale to which geometry can be defined to also occur in
generalized geometries.
 It may be noted that there indications from various other physical theories that any
 theory of
quantum gravity should come naturally equipped with
a minimum length scale of the order of Planck scale.
It is also the existence of a minimum length scale in loop quantum gravity that turns the big bang into
a big bounce  \cite{z1}.
It may be noted that
  the energy required to probe a region of spacetime below Planck scale
is more than the energy required to create a mini black hole in that region of spacetime.
So, even the physics of black holes can be used to argue for the existence of  such a minimum measurable length scale
\cite{z4,z5}.

A minimum length scale is expected to occur in most theories of quantum gravity, and this can be used as a motivation
for the
generalized uncertainty principle (GUP)
\cite{z2,zasaqsw,csdcas,cscds,2z,14,17,18,51,54}.
This also implies a modification of the Heisenberg algebra, and a subsequent modifications
of the coordinate representations of the momentum and Hamiltonian operators. The latter gives rise to
corrections to all quantum mechanical systems, even low energy ones.
A different kind of deformation of the Heisenberg algebra occurs in a theory called the doubly
special relativity (DSR) \cite{2,21,3}.
Apart from the velocity of light, the Planck energy is also a universal constant in DSR. The existence of this
maximum energy scale is motivated by the modification of the usual energy-momentum relation which occurs in
discrete spacetime \cite{1q}, spontaneous symmetry breaking of Lorentz invariance in string
field theory \cite{2q}, spacetime foam models \cite{3q}, spin-network in loop quantum gravity \cite{4q},
non-commutative geometry \cite{5q},
and Horava-Lifshitz gravity \cite{6q}. It may be noted that the DSR has been generalized to curved spacetime,
and the resultant theory
is gravity's rainbow \cite{n1,n2}.
It is possible to combine the deformation occurring due to GUP with the deformation occurring due to DSR
into a single deformation of the Heisenberg algebra \cite{main1,main2}.
Apart from the one dimensional case, this deformed  Heisenberg algebra gives rise to non-local fractional derivative terms.
However, it has been possible to deal with such terms in the framework of
harmonic extension of functions \cite{hes1, hes2, 6a, 7a, 8a, 9a}. One of the most interesting consequences of this
algebra is that  it predicts a discrete structure for space \cite{main1}.
In fact, it has been possible to obtain similar results
by using the  modified Dirac equation as well \cite{main2}.
The deformed Poisson bracket consistent with this deformed Heisenberg algebra also turns
the big bang into a big bounce \cite{z6}.

All this work has been done by deforming the first quantized commutation relations.
However, it is also possible to modify the second
quantized commutation relations. In fact, a deformed Wheeler-DeWitt equation has been constructed by
using a deformation of the second quantized commutation relations \cite{wh, wh12}. In this deformed
Wheeler-DeWitt equation the big bang singularity is naturally avoided, as there is a minimum value for the
scaling factor of the universe. So, in this paper, we will also analyse such  deformation of the  Wheeler-DeWitt equation for the
generalized geometry. We will also use this deformed Wheeler-DeWitt equation for proposing a solution to the problem of time.
It may be noted that the modification of the quantum cosmology by loop quantum gravity,
has led to the development of loop quantum cosmology, and it has been demonstrated that
the    big bang and big crunch singularities can be resolved in this formalism \cite{loaop, loaop2}.
This is because  the quantum Hamiltonian
constraint of loop quantum cosmology  does not break down at the point
where the classical singularity occurs \cite{loap4, loap5}.
The Wheeler-DeWitt equation for a universe in the brane world model has been analysed
using the mini superspace formalism, and it has been demonstrated that this cosmological model is non-singular
\cite{brane2}. The Wheeler-DeWitt equation has also been used  for studding a
quantum version of the classical repulsive phase of gravity, and it was observed that this repulsive phase of gravity
can produce a non-singular universe \cite{repulsivea}. In this model,   a free scalar field is minimally coupled to gravity
with the wrong sign, and this produces the repulsive phase of gravity.
The Wheeler-DeWitt equation has also been used for analyzing a low-energy  string cosmological,
and it was observed that in this cosmological model there exists
a finite quantum transition probability between two cosmological
phases, even when classically they are  separated by a curvature singularity \cite{string12cosmo}.
The Wheeler-DeWitt has been used for
quantization of  a homogeneous and isotropic universes with a perfect fluid which is
described by a generalized Chaplygin gas \cite{universee}.
It was demonstrated that for this model, the singularities can be  avoided if the wave function vanishes
in the region of the classical singularity. A ultraviolet   modification of gravity called the gravity's rainbow has also
been used to remove the singularities \cite{rainbow, rainbow2}. The Wheeler-DeWitt equation for
gravity's rainbow has also been studied, and this deformation of gravity  has been used to    remove
the singularity \cite{rain, rain2}.  Thus, there are various different modifications of the Wheeler-DeWitt equation
and quantum cosmology which can be used to avoid the singularity in the universe. As the singularity can also be removed
using a deformation of the Wheeler-DeWitt equation motivated from the generalized uncertainty principle \cite{wh, wh12},
we will use this deformed Wheeler-Dewitt equation equation in this paper. It will be demonstrated that this deformation
of the Wheeler-DeWitt equation can be used to solve the problem of time. This deformation of the Wheeler-DeWitt equation
will be motivated from a deformation of the generalized geometry by the extended structure of M2-branes.

 \section{Wheeler-DeWitt Equation for the Generalized Geometry}

The bosonic part of the  action for  M-theory can be written in terms of an abelian three form gauge field
and eleven dimensional gravity.
The action for gravity   can be written using
   the Ricci scalar $R$ for the eleven dimensional metric $g_{ab}$.
The action for the three form gauge field can be written as a sum of  the standard kinetic term
 and a Chern-Simons like term. These terms are constructed using a three form abelian gauge field $C_{abc}$.
The field strength $F_{abcd}$   for this   three form abelian gauge field  $C_{abc}$ is
$ F_{abcd} = 4\partial_{[a} C_{bcd]}
$.
Now the action for the bosonic part of the M-theory
\begin{eqnarray}
I &=&  \int_{\cal M} d^{11} x \sqrt{g} R -\int_{\cal M} d^{11}x\,\sqrt{g}\ \frac{1}{48}F_{abcd}F^{abcd} \nonumber \\ && +
\lambda \int_{\cal M} d^dx \ \eta^{abcdefghijk}C_{abc}F_{defg}F_{hijk},
\end{eqnarray}
where $g=-\det{g_{ab}}$ and $\eta$ is the eleven-dimensional alternating tensor density.
It may be noted that
  supersymmetry can be used to fix the value of  $\lambda$ to be $2^{-7}3^{-4}$. The  standard kinetic term
  is invariant under the following gauge transformations,
$ C_{abc} \rightarrow C_{abc} + 3\partial_{[a} \Lambda_{bc]}
$.
This  Chern-Simons term is
 gauge invariant only up to boundary terms.
The spacetime metric $g_{ab}$ can be written in terms of  a lapse function
$\alpha$ and a shift vector $\beta^i$, if the topology of spacetime is such that it can be
foliated by a family of surfaces of constant time.
 Now using this decomposition, the   gravitational
action can be written as
\begin{eqnarray}
I_{grav}= \int_{\cal M} d^{10} x\, dt\ \alpha\gamma^{\frac{1}{2}}
\bigl(R(\gamma)
+K_{ij}K^{ij} - K^2\bigr).
\end{eqnarray}
Here
$R(\gamma)$ is the Ricci scalar
formed from the spatial metric $\gamma$,  $K$, the trace of the second
fundamental form $K=K_{ij}\gamma^{ij}$, where
$ K_{ij}= (D_i\beta_j + D_j\beta_i - \dot\gamma_{ij} )/{2\alpha} .
$ Here $D$ is the covariant derivative operator formed from
$\gamma_{ij}$, and dot denotes  differentiation with respect to $t$. The    Hamiltonian constraint for the eleven dimensional
gravity can be written as
\begin{eqnarray}
{ H} = \gamma^{-1/2}\bigl(\pi^{ij}\pi_{ij} - \frac{1}{9}\pi^2 -
\gamma R\bigr),
\end{eqnarray}
where canonical momenta $\pi^{ij}$  is conjugate  to $\gamma_{ij}$. The kinetic term for this Hamiltonian constraint can be written
$   {G}_{ijkl}\pi^{ij}\pi^{kl}
$.
It is also possible to define ${{G}}^{ijkl}$  using
\begin{eqnarray}
{G}_{ijkl}{{G}}^{klmn} = \frac{1}{2}(\delta_i^m\delta_j^n
+\delta_i^n\delta_j^m),
\end{eqnarray}
So, we can write
\begin{eqnarray}
{{G}}^{klmn}= \frac{1}{2}\gamma^{1/2}\biggl(\gamma^{km}\gamma^{ln}
+\gamma^{kn}\gamma^{lm} - 2\gamma^{kl}\gamma^{mn}\biggr).
\end{eqnarray}
Now the norm of any  symmetric tensor $h_{ij}$ at any point in space can be written as
\begin{eqnarray}
|| h ||^2 = {{G}}^{ijkl}h_{ij}h_{kl}.
\end{eqnarray}
This will become important when we deform the geometry.
It may be noted that at
  the quantum level the  classical
  constraints ${H}=0$, becomes the Wheeler-DeWitt equation,
  \begin{equation}
   \mathcal{H} \psi [\gamma]=0.
  \end{equation}
We chose natural units and set $\hbar = c =1$.

The abelian three form gauge field   $C_{abc}$ can also be written in terms of  its purely spatial
components  $C_{ijk}$ and the components having a temporal part $C_{0ij}=B_{ij}$.
It may be noted that the field string of $C_{ijk}$ can be written as
$F_{ijkl}= 4 \partial_{[i} C_{jkl]}
$, and the field strength of $B_{ij}$  can be written as
$G_{ijk} = 3\partial_{[i} B_{jk]}
$. So,  the action for the three-form theory can be written as
\begin{eqnarray}
I_C  &=& \int_{\cal M }d^{10}x\,dt\,\alpha \gamma^{1/2}\ -\frac{1}{48}
\Bigl(F_{ijkl}F^{ijkl} - \frac{4}{\alpha^2}
\beta^i\beta^mF_{ijkl}F_m{}^{jkl}\nonumber \\&& +\frac{8}{\alpha^2}\dot C_{ijk}G_{ijk}
-\frac{4}{\alpha^2}G_{ijk}G^{ijk}
+ \frac{8}{\alpha^2}\beta^l\dot C_{ijk}F_l{}^{ijk}
- \frac{8}{\alpha^2}\beta^lG_{ijk}F_l{}^{ijk}\Bigr)\nonumber \\ &&
 +\lambda\eta^{ijklmnpqrs}\Bigl(3B_{ij}F_{klmn}F_{pqrs}  -
8C_{ijk}\dot C_{lmn}F_{pqrs}\nonumber \\ &&
+8C_{ijk}G_{lmn}F_{pqrs} \Bigr)  - \frac{4}{\alpha^2}\dot C_{ijk}\dot C_{lmn}
\gamma^{il}\gamma^{jm}\gamma^{kn}
\end{eqnarray}
Let    canonical momenta $\pi_B^{ij}$ be conjugate
to $B_{ij}$ and the canonical momenta $\pi^{ijk}$ be conjugate to $C_{ijk}$.
The Hamiltonian constraint  for the bosonic part of M-theory  can  be written as \cite{geha}
\begin{eqnarray}
  H = \gamma^{-1/2}\Bigl(\pi^{ij}\pi_{ij}-\frac{1}{9 }\pi^2
+3\pi^{ijk}\pi_{ijk} + \gamma(-R + \frac{1}{48}F^{ijkl}F_{ijkl})\Bigr).
\end{eqnarray}

Just as strings act as probes in string theory, the  M2-branes act as probes in M-theory. The bosonic part of the M2-brane
action is
\begin{eqnarray}
\tilde I= \int d^3\xi \sqrt{-h}
\left(\frac{1}{2}h^{\mu\nu}\partial_\mu X^a
\partial_\nu X^b g_{ab}  +
  \frac{1}{6}\epsilon^{\mu\nu\rho}\partial_\mu X^a
\partial_\nu X^b \partial_\rho X^c C_{abc}- \frac{1}{2}\right),
\label{eq:M2action}
\end{eqnarray}
where $h=\det{h_{\mu\nu}}$ and $\epsilon^{\mu \nu \rho}$ denotes the alternating tensor.
If we  compactify this theory, such that there are
$d$ commuting Killing vectors, then the   metric and three-form
fields will be independent of  $X^a$ associated with these Killing vectors. So,
the equation of motion    can be written as
$ h_{\mu\nu} = \partial_\mu X^a \partial_\nu X^b g_{ab} \label{eq:metric}
$ and $
\partial_\mu \mathcal{G}^\mu_a =0$, where we have $
\mathcal{G}^\mu_a = \bigl(\sqrt{-h}g_{ab}\mathcal{F}_\mu^b + \frac{1}{\sqrt{2}}
C_{abc}\tilde \mathcal{G}^{\mu\ bc}\bigr)
$, $
\mathcal{F}_\mu^a = \partial_\mu X^a, \label{eq:FS} $
and $
\tilde \mathcal{G}^{\mu\ ab}=\frac{1}{\sqrt{2}}\sqrt{-h}\epsilon^{\mu\nu\rho}
\mathcal{F}_\nu^a\mathcal{F}_\rho^b \label{eq:fsg} $. Furthermore, we have
$
\partial_\mu \tilde \mathcal{G}^{\mu\ ab}=0 \label{eq:BI}
$, and this   equation is   the  Bianchi identity.  In the dual description   the roles of
Bianchi identities and equations of motion gets exchanged. Motivated by the equation of motion and
Bianchi identities, it is possible to define
$
\tilde \mathcal{F}_{\mu\ ab} = \partial_\mu y_{ab}
$, and write a generalized displacement as
$ \mathcal{F} _\mu^M =  (\mathcal{F}_\mu^a,  \tilde \mathcal{F}_{\mu\ ab}) $,
and
a generalized field strength  as
$ \mathcal{G}_{\mu\ M} = (\mathcal{G}_{\mu\ a}, \mathcal{G}_\mu^{ab}) $.
The both equation of motion and Bianchi identity can then   be written as
$
\partial_\mu \mathcal{G}^\mu_M = 0 $.
The field strengths are then related to the displacements by a generalized metric, such that
$\mathcal{G}_{\mu\ M} = M_{MN}\mathcal{F}^N_\mu$, where  \cite{geha}
\begin{eqnarray}
M_{11} = g_{ab}+\frac{1}{2}C_a^{ef}C_{bef}, &&
M_{21}=  \frac{1}{\sqrt{2}}C_a^{kl},\nonumber \\
M_{12} = \frac{1}{\sqrt{2}}C^{mn}_b,   &&
M_{22} =   \frac{1}{2}(g^{mk}g^{nl}-g^{ml}g^{nk}).
 \end{eqnarray}
A generalized infinitesimal line element in this generalized geometry can be written as
\begin{eqnarray}
d{s}^2 = M_{MN}dZ^M dZ^N
\end{eqnarray}
where $Z^M=(x^a,y_{ab})$.
Thus, the transformation $y_{ab}
\rightarrow y_{ab}+\frac{1}{\sqrt{2}}\lambda_{ab}$, and the invariance of the
  generalized line element,   induces a gauge transformation of the three form field,
$C_{abc} \rightarrow C_{abc} + \partial_{[a}\lambda_{bc]}$.
Now a potential term can be defined \cite{geha}
as \begin{eqnarray}
\mathcal{V}(M) &=& \gamma^{1/2} \Biggl(\frac{1}{12} M^{MN} (\partial_M M^{KL})( \partial_N
M_{KL} ) \nonumber \\ && - {\frac{1}{2}} M^{MN} (\partial_N M^{KL}) (\partial_L M_{MK}) \\ \nonumber  &&
+\frac{1}{12}   M^{MN} (M^{KL} \partial_M M_{KL})(M^{RS} \partial_N M_{RS})
\nonumber \\ && +\frac{1}{4}    M^{MN} M^{PQ}(M^{RS} \partial_P M_{RS}) (\partial_M M_{NQ})
\Biggr),
\end{eqnarray}
where $\partial_M = \bigl(\frac{\partial}{\partial x^a},
\frac{\partial}{\partial y_{ab}}\bigr)$. Let $  V[H] $ be the spatial part of this potential term. Here    the spatial
part of the metric $M$ is denoted by $H$.  This potential part for the spatial metric can be written as \cite{geha}
\begin{eqnarray}
 V = \gamma( R - \frac{1}{48}F^{ijkl}F_{ijkl}).
\end{eqnarray}
Furthermore, it is possible write the kinetic part of this equation as
\begin{eqnarray}
 K =  H^{-1/2} G^{ABCD}\pi_{AB}\pi_{CD},
\end{eqnarray}
where
\begin{eqnarray}
{{G}}^{ABCD}= \frac{1}{2}H^{1/2}\biggl(H^{AC}H^{BC}
+H^{AD}H^{BC} - 2H^{AB}H^{CD}\biggr).
\end{eqnarray}
Using this metric the kinetic part becomes
\begin{eqnarray}
 K =  \left(\pi^{ij}\pi_{ij}-\frac{1}{9 }\pi^2
+3\pi^{ijk}\pi_{ijk} \right).
\end{eqnarray}
So, now the Hamiltonian constraint for the bosonic part of the M-theory, can be written in terms of the generalized geometry as
\begin{equation}
 H =  K + V .
\end{equation}
This Hamiltonian constraint gives rise to the Wheeler-DeWitt equation for the generalized geometry,
and this equation acts  on the wave function defined on the space of generalized metrics $H$.
\begin{equation}
 \mathcal{H} \psi[H] =0.
\end{equation}
This can be  done by promoting the generalized metric to an operator and using
 \begin{eqnarray}
 [H_{AB} (Z_1), \pi^{CD} (Z_2)] &=& i
 (\delta_A^C\delta_B^D + \delta_A^D \delta_B^C) \delta( Z_1- Z_2 ).
\end{eqnarray}
This corresponds to writing
\begin{eqnarray}
 \pi^{AB} = -i \frac{\delta}{\delta H_{AB}}.
\end{eqnarray}
So, the Wheeler-DeWitt equation for the generalized geometry can be explicitly written as
\begin{eqnarray}
 -  H^{-1/2}   G^{ABCD}\frac{\delta }{\delta H^{AB}}\frac{\delta}{\delta H^{CD}}\psi [ H] + V[H] \psi [H] =0.
\end{eqnarray}
This  equal to the usual Wheeler-DeWitt equation for a gravity and an abelian three form field, if this generalized
metric will be expanded in terms of  the usual metric and the three form field,
\begin{eqnarray}
 \mathcal{H} \psi [H] &=&
 -  H^{-1/2}   G^{ABCD}\frac{\delta }{\delta H^{AB}}\frac{\delta}{\delta H^{CD}}\psi [ H] + V[H] \psi [H] \nonumber \\ &=&
 \gamma^{-1/2}\Biggl(- G^{ijkl}\frac{\delta}{\delta \gamma_{ij}}\frac{\delta}{\delta \gamma_{kl}}
 + \frac{1}{9 }G^{ijij}\frac{\delta}{\delta \gamma_{ij}}\frac{\delta}{\delta \gamma_{ij}}\nonumber \\&&
-3\frac{\delta}{\delta C^{ijk}}\frac{\delta}{\delta C_{ijk}} \Biggr)\psi[\gamma, C]\nonumber \\&&
+ \gamma^{1/2}\left( R - \frac{1}{48}F^{ijkl}F_{ijkl}\right)\psi[\gamma, C]
 \nonumber \\ &=&0.
\end{eqnarray}
Hence, we have derived the Wheeler-DeWitt equation
for the generalized geometry. However, we did not analyse the effect that the extended structure of M2-branes will
have on this generalized geometry. In the next section, we will demonstrate that the extended structure of M2-branes will deform this Wheeler-DeWitt
equation.

\section{ Deformed Wheeler-DeWitt Equation}
In this section,  we will analyse the effect that the extended structure of M2-branes will have on the Wheeler-DeWitt equation
for the generalized geometry. It is known that M2-branes have an extended structure, and this limited the
extent to which the generalized
geometry can be probed. Now we can define a norm for the any symmetric tensor  in generalized geometry as follows,
any  symmetric tensor $A_{AB}$ at any point in space can be written as
\begin{eqnarray}
|| A ||^2 = {{G}}^{ABCD}A_{AB}A_{CD}.
\end{eqnarray}
We can use this definition of norm to say that we should not be able to probe the geometry below a certain resolution
say $\Delta A$, as M2-branes have an extended structure. However, according to the second commutation relations we can
probe the $H_{AB}$ to arbitrary accuracy as long as we do not probe $\pi^{AB}$. We now want to deform these commutation relations such that
we will only be able to probe $H$ to some mini measurable scale  fixed by the extended structure of M2-branes,
$\Delta H \geq \Delta H_{min}$. Here $\Delta H_{min}$ is fixed by the extended structure of M2-branes.
This situation is similar to the occurrence of
 a minimum length in  first quantized system, $
 \Delta x \geq \Delta x_{min} \geq \alpha_0 \ell_{Pl}$. So, we will now formally generalized this deformed first quantized algebra, to infinite degrees
 of freedom, and thus obtain a formal expression for the deformation of the Wheeler-DeWitt equation in generalized geometry.

Such a deformation of the first quantized algebra, deformed by
both GUP and DSR is given by
\begin{eqnarray}
 [x^i, p_j] &=& i \left[  \delta_{j}^i - \alpha ||p|| \delta_{j}^i + \alpha ||p||^{-1} p^i p_j \right.
   \nonumber \\   && \left.
  + \alpha^2 p^2 \delta_{j}^i + 3 \alpha^2 p^i p_j\right],
\end{eqnarray}
where
\begin{equation}
 || p || = \sqrt{ p^i p_i},
\end{equation}
and  $\alpha = {\alpha_0}/{M_{Pl}c} = {\alpha_0 \ell_{Pl}}/{\hbar}$
is the parameter measuring the strength of this deformation.
%It is expected to  contribute only below the electroweak scale.
In the one dimensional case this
corresponds to the uncertainty relation given by $\Delta x \Delta p =
[1 - 2 \alpha <p> + 4 \alpha^2 <p^2> ]$, and  which in turn   imply the existence of a minimum length
$
 \Delta x \geq \Delta x_{min} \geq \alpha_0 \ell_{Pl}$, and a
 a maximum
momentum
$
 \Delta p \leq \Delta p_{max} \leq \alpha_0^{-1} M_{Pl} c
$.
Now the momentum in the coordinate representation can be written as
\begin{equation}
 p_i = \tilde p_i (1 - \alpha ||\tilde p||  + 2\alpha^2 ||\tilde p||^2),
\end{equation}
where $ \tilde p_i = - \partial_i$ is the coordinate representation of the original  momentum.

It may be noted that there exists the   soccer ball problem with such a deformation of the usual
energy-momentum relation \cite{s, s1}. This can effect the physics of branes   spiting to form other branes, and other
process which can bring about topological change in the generalized geometry.
However, as the solution for the Wheeler-DeWitt equation  in the full superspace are only  formally studied,
we can formally define such a deformation for   field theories. This can be done by formally taking
 the continuum limit $N \to \infty$, where $N$ is the degrees of freedom of the system.
 We would like to stress the fact    that such a limit is only a formal limit, and can be used to analyse
 a system explicitly only in the minisuperspace approximation.
 However, we can obtain information from such a deformation even in the full superspace.
 This is because we can    formally argue that
  a deformed Wheeler-DeWitt equation should be free from singularities,  in the full superspace.  It may be noted as
  it is only possible to    analyse the solutions of the full functional Wheeler-DeWitt equation formally, we do not lose
  much by incorporating such a deformation. However, the advantage of using such a deformation is that there is a limit
  on the resolution of fields, and hence, this deformed Wheeler-DeWitt equation is free from any singularities. This can be physically
  relevant for pure gravity, when we the three form field vanishes. This can have direct relevance to cosmology, if
    we   analyse the deformed Wheeler-DeWitt equation explicitly   in the  minisuperspace approximation.
  In the minisuperspace approximation, the second quantized version of the soccer ball problem becomes well defined. This is because in the
  multiverse, we can define  $\pi_1$ as the momentum density conjugate scalar factor of a universe, and $\pi_2$
  as the the momentum density conjugate scalar
  factor of another  universe. Then it is possible to analyse the collision of these two universes using the formalism of
 ekpyrotic universes  \cite{2e, 1e, e2, e1}.  Now
if this system is analysed using the  deformed second quantized algebra,  and $\pi$ is total momentum density  of these two universes after collision,
then $\pi \neq \pi_1 + \pi_2$. This is the second quantized version of the soccer ball problem, which will occur in the
    ekpyrotic universes  \cite{2e, 1e, e2, e1}. Hence, there ekpyrotic universes  cannot be studied in this deformed
  Wheeler-DeWitt equation.
  The soccer ball problem will also occur for the creation and annihilation of two   black holes \cite{b, b1}
  or  a virtual black hole loop \cite{v, v1}.
  So, creation and annihilation of     black holes,   or the virtual black hole loop    cannot  be studied
  using this deformed Wheeler-DeWitt equation,  even in the minisuperspace approximation.
  We will not be dealing with any of these issues, and we will be only analyzing  a single universe in the
   minisuperspace approximation.  Furthermore, as we only want to analyse the effect of such a deformation on
   generalized geometry,  we will
  formally   take the
 continuum limit $N \to \infty$,  and write
\begin{equation}
 [\phi (x ), \pi (y)] = i \delta( x- y ) + i \alpha \mathcal{A} (x, y)  + i \alpha^2 \mathcal{B} (x, y),
\end{equation}
where
\begin{eqnarray}
 \mathcal{A} (x, y) &=& || \pi||\delta(x-y) + ||\pi||^{-1} \pi (x) \pi (y),  \nonumber \\
 \mathcal{B} (x, y) &=& ||\pi||^2 \delta(x-y) + 3 \pi(x) \pi (y).
\end{eqnarray}
Here the definition of  the norm of $||\pi||$   has to be suitably normalized. It might be possible to
define the norm in a better way by smearing it with a distribution on a compact support.
This deformed algebra corresponds to taking the following  deformation for $\pi (x) $,
\begin{equation}
 \pi (x) =  \left( 1 -  \alpha   || \tilde \pi || + 2 \alpha^2 || \tilde \pi||^2 \right)
\tilde \pi (x).
\end{equation}
where
\begin{equation}
  [\phi (x ), \tilde \pi (y)] = i \delta( x- y ).
\end{equation}
Thus, $\pi (x)$ is the usual momentum density  conjugate to $\phi (x)$, and it can be written in the Wheeler-DeWitt approach as,
\begin{equation}
\tilde \pi (x) = -i \frac{\delta }{\delta \phi (x)}.
\end{equation}

We can perform this deformation for every field, and so, we will now review this deformation for
the generalized geometry.
 The Hamiltonian constraint  of the generalized geometry
 becomes the Wheeler-DeWitt equation for the generalized geometry,
  \begin{equation}
   \mathcal{H} \psi [H]=0.
  \end{equation}
  This is usually done by imposing standard commutation relations.
  However, a deformed version of the Wheeler-DeWitt equation can be  constructed
  by using a deformed version of the commutation relations
 \begin{eqnarray}
 [H_{AB} (Z_1), \pi^{CD} (Z_2)] &=&
 (\delta_A^C\delta_B^D + \delta_A^D \delta_B^C) [i \delta( Z_1- Z_2 )  \nonumber \\ && + i \alpha \mathcal{A} (Z_1, Z_2)
 + i \alpha^2 \mathcal{B} (Z_1, Z_2)],
\end{eqnarray}
where
\begin{eqnarray}
 \mathcal{A} (Z_1, Z_2) &=& || \pi||\delta(Z_1-Z_2) + ||\pi||^{-1} {{G}}^{ABCD} \pi_{AB} (Z_1) \pi_{CD} (Z_2),  \nonumber \\
 \mathcal{B} (Z_1, Z_2) &=& ||\pi||^2 \delta(Z_1-Z_2) + 3 {{G}}^{ABCD} \pi_{AB}(Z_1) \pi_{CD} (Z_2).
\end{eqnarray}
Here the norm is defined as
\begin{eqnarray}
|| \pi || = \sqrt{{{G}}^{ABCD}\pi_{AB}\pi_{CD}}.
\end{eqnarray}
 This corresponds to  the following deformation of $\pi (x) $
\begin{equation}
 \pi_{AB} (Z) = \left( 1 -  \alpha   || \tilde \pi || + 2 \alpha^2 || \tilde \pi||^2 \right)
\tilde \pi_{AB} (Z).
\end{equation}
where
\begin{equation}
 [H_{AB} (Z_1), \tilde \pi^{CD} (Z_2)]  = i (\delta_A^C\delta_B^D + \delta_A^D \delta_B^C) \delta( Z_1- Z_2).
 \end{equation}
Thus, $\pi^{CD}$ is the usual momentum density conjugate to $H_{AB}$,
and it can be written in the Wheeler-DeWitt approach as,
\begin{equation}
\tilde \pi^{AB} (Z) = -i \frac{\delta }{\delta H_{AB} (Z)}.
\end{equation}
We have now deformed the second quantized system, and
 so this deformation will limit the resolution of generalized geometry  to $ \Delta H \geq \Delta H_{min} $, and hence the fields
 cannot be resolved below this limit, which is fixed by the extended structure of the M2-branes, which are used
 as probes in generalized geometry.

 Now we can neglect the effect coming from the three from field and approximate this system using general relativity.
 If we also neglect the effect of compactification from eleven dimensions to four dimensions, we can write th deformed Wheeler-DeWitt equation for
 the general relativity as
 \begin{eqnarray}
  \mathcal{H} \psi [\gamma] =0,
 \end{eqnarray}
where $\gamma$ is the metric in our universe. In this expression, we can used th following deformation,
 \begin{eqnarray}
 [\gamma_{ij} (x), \pi^{kl} (y)] &=&
 (\delta_i^k\delta_j^l + \delta_i^l \delta_j^k) [i \delta( x- y )  \nonumber \\ && + i \alpha \mathcal{A} (x, y)
  + i \alpha^2 \mathcal{B} (x-y)],
\end{eqnarray}
where
\begin{eqnarray}
 \mathcal{A} (x, y) &=& || \pi||\delta(x-y) + ||\pi||^{-1} G^{ijkl} \pi_{ij} (x) \pi_{kl} (y),  \nonumber \\
 \mathcal{B} (x, y) &=& ||\pi||^2 \delta(x-y) + 3G^{ijkl} \pi_{ij} (x) \pi_{kl} (y).
\end{eqnarray}
 This again corresponds to  a  deformation of $\pi (x) $ by higher functional derivative terms.
 It may be noted that this deformation
  will prevent the occurrence of singularities in the deformed Wheeler-DeWitt equation.  Hence, we can
 formally argue that in the deformed Wheeler-DeWitt equation for gravity, there is a
 minimum value to the resolution of the metric,
  $ \Delta \gamma \geq \Delta
  \gamma_{min} $,
and so this equation should not have  singularities. As we can only  analyse the Wheeler-DeWitt equation
formally in full superspace model,
  we  will use minisuperspace approximation to
analyse this non-singular   Wheeler-DeWitt equation explicitly, in the next section.

\section{Problem of Time}

In this section, we will analyze
a minisuperspace approximation to the  deformed Wheeler-DeWitt equation.
We will analyse a simple model to demonstrate how the problem of time can be solved using
such a deformed Wheeler-DeWitt equation, and this formalism can be applied for analyzing the deformation
of any cosmological model.
We will consider the universe to be filled with a vacuum   energy density, and this
dominates the  universe during inflation and during the late time evolution of the universe.
However, we will also include the contribution coming from
from radiation, as the universe after inflation was dominated by both radiation and vacuum energy.
As these are the most important phases in the evolution of the universe, we will analyse a universe
filled with vacuum energy and radiation.
Furthermore, in this paper, we want to demonstrate how the deformation of the Wheeler-DeWitt equation
can be used to solve the problem of time, and the Wheeler-DeWitt equation for such a universe is
very simple, and so we will analyse this cosmological model in this paper.
Now consider a
closed universe filled with a vacuum of constant
energy density and the radiation, $\rho(a)=\rho_v+ \epsilon/a^4,$ where $\rho_v$ is the
vacuum energy density,  $\epsilon$ is a
constant characterizing the amount of radiation, and  $a$ is the scale factor.
The details properties of $\rho(a)$ will not be important for the main result of this paper,
and we could have used a slightly different kind of cosmological model for this analysis.
We can write the
Friedman-Robertson-Walker metric   as follows,
\begin{equation}
ds^{2}=-N^{2}dt^{2}+a^{2}\left(  t\right)  d\Omega_{3}^{2},
\end{equation}
where $d\Omega_{3}^{2}$ is the  line element on the three sphere.
It may be noted that we have considered $K = 1$, and a similar solution could be obtained for $K = 0, -1$,
however, the main aim of this paper
is to demonstrate how the problem of time can be solved for a simple cosmological model, so we take $K=1$.
Using this metric, the  Lagrangian for this closed universe, in the minisuperspace approximation.
can be written as   \cite{ab},
\begin{equation}
  \mathcal{L}=-\frac{3\pi}{4G}a\dot{a}^2+\frac{3\pi a}{4G}-2\pi^2a^3\rho(a),
\end{equation}
The equation of motion  obtained from this Lagrangian, can be written as
\begin{equation}
\label{Fri}-\frac{3\pi}{4G}a\dot{a}^2-\frac{3\pi}{4G}a+2\pi^2a^3\rho(a)=0.
\end{equation}
We can also calculate the Hamiltonian  from this Lagrangian,
\begin{equation}
H =-\frac{G}{3\pi}\frac{\pi^2}{a}-\frac{3\pi}{4G}a+2\pi^2a^3\rho(a).
\end{equation}
Now we impose a deformed commutation relation for the momentum operator, and hence, write the
 deformed momentum operator for this minisuperspace model as
 \begin{equation}
\pi = \tilde \pi  (1 - \alpha ||\tilde \pi ||  + 2\alpha^2 ||\tilde \pi ||^2),
\end{equation}
where $ \tilde \pi  = - i da /da$. Therefore we get
\begin{equation}
\pi = -i\left(1 + i \alpha \frac{d}{da} - 2\alpha^2 \frac{d^2}{d^2a}\right)\frac{d}{da}.
\end{equation}
Now we can write the deformed
Wheeler-DeWitt equation as follows,
\begin{equation}
 \frac{G}{3\pi} \frac{d^2 \psi (a)}{ d a^2 } - 2 \alpha i \frac{G}{3 \pi} \frac{d^3 \psi (a)}{ d a^3} + M^2 (a)\psi  =0,
\end{equation}
where $M^2 (a) = 3\pi a^2/ 4 G + 2 \pi ^2 a^4 \rho (a) $ and $k ^2 (a) = 3\pi M^2 (a)/ G $.
It may be noted that it is important to study the stability of such a system, however, we assume the coefficient of the
deformation to be small enough not to make the system unstable. Furthermore, there seems to be no problem with the solution
of the Wheeler-DeWitt equation involving higher derivative corrections \cite{hd}.
Hence, we will analyse the consequences of the solution of this
deformed Wheeler-DeWitt
equation.
It may be noted that this  modified quantization is consistent with the following uncertainty relation,
\begin{equation}
 \Delta a \Delta \pi = 1 - 2 \alpha <\pi > + 4 \alpha^2 <\pi^2> .
\end{equation}
 This  implies the existence of a minimum scale factor for the universe
$
 \Delta a \geq \Delta a_{min} $, and so, the big bang singularity is naturally avoided in this model \cite{wh}.
 It may be noted that even though we were able to formally argue that the deformation of the Wheeler-DeWitt equation would eliminate singularities, we
 have now demonstrated this explicitly in the minisuperspace approximation.
Now assuming $\psi = \exp (ma)$ we get
\begin{equation}
 m^2 - 2i  \alpha m^3 + k^2 (a) = 0.
\end{equation}
The solution to the leading order in $\alpha$, can be written as $m = \{ i k', -i k'', i/2 \alpha\}$,
where $k' = k (1 - k \alpha)$, and $k'' = k (1 +  k  \alpha)$.
So, we can write the solution for this equation as
\begin{equation}
 \psi = A e^{i k'a} + B e^{-ik''a} + C e ^{-a/2 \alpha},
\end{equation}

Even though the date from type I supernova indicates
that our universe is an accelerating in its expansion
\cite{supera,super1111,super2222,super3,super4,super5}, there are predictions from
string theory that the era of rapid expansion might only be the first era in the evolution of our universe
\cite{string, string1, string2, string4}.
This is also consistent with predictions from loop quantum gravity \cite{loop, loop1}, where
it is shown that the era of accelerated expansion will be followed by an era of
contraction.
It may be noted that unstable geometries can mathematically
decay into bubble of nothing, which contain neither matter nor spacetime \cite{1w, 2w}.
This mathematical structure where neither matter nor spacetime is present can be viewed as a
third quantized vacuum state  \cite{thir1, thir2, thirvac2, thirvac4}.
We can also add non-linear terms to the Wheeler-DeWitt equation, and in that case
the universes can get created and annihilated in the third quantized formalism, just as the particles get created
and annihilated in the second quantized formalism \cite{thir, thirvac, thirvac1, thir5, th}.
Even though we  cannot use the deformed Wheeler-DeWitt equation for analyzing the full process
of the creation and annihilation of the universes in the multiverse, but we can use this formalism to motivate
the boundary conditions for our universe, in the second quantized formalism.
Thus, motivated from all these models, we can impose the following boundary conditions on the
wave function of the universe, $\psi (a_0) = \psi (a_0 + \delta a) =0$.
This corresponded to the creation of the wave function of the universe from nothing and its subsequent annihilation into nothing.
Another possibility for this boundary condition is the formation the universe because of a tunneling process \cite{tunn, tunn1}.
Furthermore, the universe could be formed around a metastable vacuum state, and hence can tunnel to a true vacuum state \cite{meta, meta1}.
This will correspond to a spontaneous
annihilation of all structure in the universe, and hence justify our boundary conditions.
So, implementing the boundary conditions for the initial state of the universe, we obtain,
\begin{eqnarray}
 A e^{i k'a_0} + B e^{-ik''a_0} + C e ^{-a/2 \alpha_0} &=& 0, \nonumber \\
  A e^{i k'(a_0}e^{ \delta a )} + B e^{-ik'' (a_0 }e^{ \delta a )} && \nonumber \\  + C e ^{-a/2  (a_0 } e ^{ \delta a )} &=& 0.
\end{eqnarray}
Now we let $C =0$, so,   we can write
\begin{eqnarray}
 A+ B + C =0, \nonumber \\
 A e^{k' \delta a} + B e^{- k'' \delta a} + C e^{- \delta a /2 \alpha} =0.
\end{eqnarray}
Thus, we obtain the following result,
\begin{eqnarray}
 \psi &=& 2 i A \sin (ka) + C [ e ^{-ka} + e ^{- k/2\alpha} ] \nonumber \\ && + \alpha k^2 a [ iC e^{ika} + 2 A \sin( ka)].
\end{eqnarray}
Next implementing the boundary conditions for the final state of the universe,  we obtain,
\begin{eqnarray}
 2 i A \sin ( k \delta a) &=& |C| [e^{- k \delta a + \theta_c} - e^{i (\delta a /2 \alpha  - \theta_c)  }]
 \nonumber \\ &&
 - \alpha k^2 \delta a \Big[i |C| e^{-i (k\delta a + \theta_c)}\nonumber \\ &&  + 2  A \sin(k\delta a)\Big],
  \\
  \cos (\delta a /2 \alpha  + \theta_c) &=& \cos (k \delta a + \theta_c) \nonumber \\ &=&  \cos (n \pi + \theta_c + \epsilon),
\end{eqnarray}
from which it is easy to show that
\begin{equation}
 \delta a = 2 n \pi \alpha.
\end{equation}

This shows that  the universe evolves by taking discrete jumps rather than in an continuous manner.
Thus, in this model of the universe, there exist
  finite expanding bubbles,  each bubble representing a point in the
  evolution of the universe. These bubbles   appear  and disappear  after some $\delta a$,
  thus effectively give time a discrete structure. It may be noted that we have not demonstrated
  that the fundamental structure of space or time has to be discrete in quantum gravity, but that
  the effective cosmological time has to be discrete, and this happens due to the appearance and
disappearance of finite bubbles representing a point in the evolution of the universe.
However, this can be effectively used to solve the problem of time in cosmology.
This is because  the general relativity is time re-parametrization invariant, and  so the
time translation are similar to gauge transformation. Thus, it is not possible to
  distinguish between the past and the future wave functions of the universe, and this is the
  problem of time in quantum gravity, which also occurs in cosmology
    \cite{pt01, pt02}.   It may be noted that there have been various
 proposals made to define an  intrinsic clock in quantum gravity \cite{o, o1, o2, o4, o5}, and one of the
 most  interesting   proposals  for such an intrinsic clock is to identify the evolution  of the
  scalar factor of the universe with this intrinsic clock measuring time \cite{i, i1}.
  Thus, it is possible to use the expansion of the universe as a clock in cosmology. However, the
  problem to differentiate between different points of the evolution of the universe still exists due
  to the   time re-parametrization invariance of general relativity. So, only by breaking the time
  re-parametrization invariance of general relativity can the problem of time be solved in cosmology.
Thus,
in this paper, we have also implicitly  identify the evolution of the scalar factor with the intrinsic clock
  of the universe. However, the advantage of this present analysis is that there is a well defined notion
of such a evolution, as the universe takes discrete jumps, and the time re-parametrization invariance is broken
between such discrete jumps. So, as  the universe evolves by taking discrete jumps,
such that   the time steps are all distinct. It may be noted that
  time re-parametrization still holds between two points separated
by distance $2 \pi \alpha$.
Thus effectively, the universe acts like a crystal in time.
It may be noted that the formation of time crystals in the Wheeler-DeWitt equation
occurs due to a combination of the boundary conditions, and the deformation of the  Wheeler-DeWitt equation.
Even though we have explicitly
demonstrated the formation of time crystals
in minisuperspace approximation, the formation of such crystals can solve the problem of time
in full superspace Wheeler-DeWitt equation. This is because for any system with time re-parametrization invariance,
\begin{equation}
\mathcal{H} \psi [x(t)]  =0,
\end{equation}
where $\mathcal{H} $ is the Hamiltonian of the system. Such a situation even
occurs in condensed matter physics, where the system
is time re-parametrization invariant. However, for such condensed matter systems,
 the time crystals have been  used to break the
time re-parametrization invariance, just as
spatial translation is broken in regular crystals \cite{time,  time1, 0time, 1time, ti}.
This gives rise to an effective time $t = \sum_i \Delta t_i$ for such systems. So,
 this system evolves by taking discrete jumps $\Delta t_i$,
 \begin{equation}
  x ( t) \to x(t + \Delta t_i).
 \end{equation}
 However, for time intervals much greater than $\Delta t_i$, we can effective write
 the Schrodinger equation
\begin{equation}
 \mathcal{H}  \psi[x (t)] = i \frac{\partial \psi[x(t)]}{\partial t }.
\end{equation}
Such systems have also been studied in condensed matter physics. In fact,
such time crystals have been studied for superconducting rings which are time re-parametrization invariant,
and hence, have no explicit time evolution \cite{time,  time1, 0time, 1time, ti}.
So, before the spontaneous breaking of time  re-parametrization invariance to form time crystals,
it is not possible to define time evolution of such systems. However, the  spontaneous breaking of time  re-parametrization invariance
gives rise to an effective time evolution for these systems. Now we can repeat this process for the full superspace Wheeler-DeWitt equation.
Now due to time re-parametrization invariance, the Wheeler-DeWitt equation for general relativity can be written as
\begin{equation}
 \mathcal{H} \psi [\gamma (t)] = 0.
\end{equation}
However, if the time re-parametrization invariance is broken for
spatial geometries separated by time $\Delta t_i$, then the three geometry evolves by taking discrete jumps $\Delta t_i$,
 \begin{equation}
  \gamma ( t) \to \gamma (t + \Delta t_i).
 \end{equation}
Now for large time intervals, the Wheeler-DeWitt equation for general relativity can be effectively written as
\begin{equation}
 \mathcal{H} \psi [\gamma (t)] = i \frac{\partial \psi[\gamma (t) ]}{\partial t }.
\end{equation}
Thus,  we have shown that
the deformation of the second quantized commutation relations   lead to the formation
of   time crystals in the deformed Wheeler-DeWitt equation, and this in turn solves the problem of time.
This is because an effective time $t$ emerges in this model due to the breaking of time re-parametrization invariance.
\section{Conclusion}
In this paper, we   first derived the Wheeler-DeWitt equation for generalized geometry which occur in M-theory.
We   then observed that as M2-branes have an extended structure, and they also act as probes for this geometry, so
their extended structure will limit the resolution to which this geometry can be defined. We
thus obtained the  Wheeler-DeWitt equation,
which was deformed to incorporate the existence of such a minimum resolution scale for the generalized geometry. We also analysed such a deformation
for general relativity by neglecting the effects coming from three from field. We analysed the cosmological implications of such an deformed
Wheeler-DeWitt equation.  We analysed the solutions for this deformed Wheeler-DeWitt equation in minisuperspace approximation
for a closed universe filled with a vacuum of constant
energy density and the radiation.
The deformation of the Wheeler-DeWitt equation caused this universe to evolve by
  taking discrete jumps rather than as a continuum.
Thus, this deformation naturally gave rise to time crystals, which broke the time re-parametrization
  of the original theory. This in turn  was used to  propose a solution for  the problem of time in quantum gravity.
We also analyze the implication of this breaking of time re-parametrization for the full
  superspace Wheeler-DeWitt equation.

Then, a deformed Wheeler-DeWitt equation was constructed. It would be interesting to analyse the effect of this
 deformation of the Wheeler-DeWitt equation on various physical systems. It has been demonstrated that there is
 an explicit correspondence between   Wheeler-DeWitt equation  and covariant formalism of quantum gravity  \cite{sc}.
 It would be interesting to investigate the effect this deformation of the Wheeler-DeWitt equation will have on the
 covariant formalism of quantum gravity. The Wheeler-DeWitt equation has also been used for analyzing the quantization
 of black holes \cite{bh}. It is known that the
 phase space of gravity for non-compact spacetimes   cannot be properly defined
without taking the boundary degrees of freedom  into consideration.  So,  the   Hamiltonian formulation for the surface terms
has also been taken into account for analyzing the quantization of black holes. This formalism has  been used to analyse the thermodynamics of black holes.
It would be interesting to use the deformed Wheeler-DeWitt equation  for quantizing black holes,  and discuss the effect of such
a deformation of the Wheeler-DeWitt equation on the thermodynamics of black holes.
The results of this paper can be used to motivate a GUP like modification for time.
This will in turn deform
  the Hamiltonian for all quantum mechanical systems. In order to do that, we will need to take time as an observable,
  e.g. with reference to the evolution of some non-stationary quantity \cite{timeq,qq, q1, q, time1q}.
It may be noted that such a deformation for the classical field equation
for   quantum field theory has been analyzed \cite{hes1, hes2}.
It would be interesting to
analyze such systems, and calculate the effect on various physical processes.
It is also possible to analyze different deformations of the
generalized  metric. In fact, a deformation of the spacetime metric has been performed using
extended conformal transformations \cite{conf}. It would be interesting
to analyze such a deformation of the generalized metric, and analyze its
consiquences.
It is possible to study  Noether symmetries in minisuperspace models, and use them to
 obtain selection rule  to recover the classical behavior \cite{crit}.
 It would be possible to study the Noether symmetries for this deformed minisuperspace model
 and use it to obtain similar selection rules.

\subsection*{Acknowledgments}
The work of SD is supported by the Natural Sciences and Engineering Research Council of Canada.
The research of AFA is supported by Benha University (www.bu.edu.eg) and CFP in Zewail City, Egypt.
MF would like to thank D. Berman for  suggesting a way to derive  the Wheeler-DeWitt equation
for the generalized geometry in M-theory.

\end{document}